\documentclass[twocolumn,showpacs,aps,prd,superscriptaddress,preprintnumbers,amsmath,amssymb,floatfix]{revtex4}

\usepackage{graphicx}
\usepackage{dcolumn}
\usepackage{bm}

\newcommand{\PUBD}  {./Definitions}  

\newcommand{\FGSD}{./Plots}

\usepackage{epsfig}
\usepackage{afterpage}
\usepackage{psfrag}
\usepackage{url}

\newcommand{\ra}{\ensuremath{\rightarrow}}

\newcommand{\mumu}{\ensuremath{\mu^{+}\mu^{-}}}

\newcommand{\nbel}{\ensuremath{{\overline{\nu}}_{e}}}

\newcommand{\GeV}{\ensuremath{\mathrm{GeV}}}
\newcommand{\MeV}{\ensuremath{\mathrm{MeV}}}

\newcommand{\nut}   {\nu_\tau}

\newcommand{\en}{\ensuremath{e \nbel}}

\newcommand{\GeVm}{\ensuremath{\mathrm{GeV/c{^2}}}}
\newcommand{\GeVp}{\ensuremath{\mathrm{GeV/c}}}

\newcommand{\ifb}{\ensuremath{\mathrm{fb^{-1}}}}

\newcommand{\bfl}{\begin{flushleft}}
\newcommand{\efl}{\end{flushleft}}
\newcommand{\bfr}{\begin{flushright}}
\newcommand{\efr}{\end{flushright}}
\newcommand{\bc}{\begin{center}}
\newcommand{\ec}{\end{center}}

\newcommand{\Like}{\ensuremath{\mathcal{L}}}
\newcommand{\Acp}{\ensuremath{\mathcal{A}_{CP}}}

\newcommand{\pizero}  {\ensuremath{\pi^{0}}}

\newcommand{\Kpm}     {\ensuremath{K^{\pm}}}
\newcommand{\pimp}    {\ensuremath{\pi^{\mp}}}

\newcommand{\K}      {\ensuremath{K}}
\newcommand{\Kstar} {\ensuremath{\K^{*}}}
\newcommand{\Kstarstar} {\ensuremath{\K^{**}}}

\newcommand{\Kstarpm} {\ensuremath{K^{*\pm}}}

\newcommand{\kstarstar} {\ensuremath{\K^{**}}}

\newcommand{\rhopm}   {\ensuremath{\rho^{\pm}}}

\newcommand{\Bpm}     {\ensuremath{B^{\pm}}}
\newcommand{\Bp}      {\ensuremath{B^{+}}}
\newcommand{\Bm}      {\ensuremath{B^{-}}}
\newcommand{\B}       {\ensuremath{B}}
\newcommand{\Bbar}    {\ensuremath{\bar{B}}}
\newcommand{\BpBm} {\ensuremath{B^{+}B^{-}}}

\newcommand{\BB}   {\ensuremath{B\bar{B}}}

\newcommand{\rhopi}   {\ensuremath{\rhopm \pizero}}
\newcommand{\Kpi}     {\ensuremath{\Kpm \pizero}}
\newcommand{\Kpipi}   {\ensuremath{\Kpm \pizero \pizero}}
\newcommand{\Kstarpi} {\ensuremath{\Kstarpm \pizero}}

\newcommand{\Btokpipi}   {\ensuremath{\Bpm \ra \Kpipi}}

\newcommand{\Btorhopi}   {\ensuremath{\Bpm \ra \rhopi}}
\newcommand{\Btokstarpi} {\ensuremath{\Bpm \ra \Kstarpi}}

\newcommand{\Kstartokpi} {\ensuremath{\Kstarpm \ra \Kpi}}

\newcommand{\Btokstarpitokpipi} {\ensuremath{\Bpm \ra (\Kstarpm \ra \Kpi)\pizero}}
\newcommand{\Btokpi} {\ensuremath{B \ra K\pi}}

\newcommand{\thetabsph} {\ensuremath{\theta_{\mathrm{Sph}}^{\mathrm{B}}}}  
\newcommand{\ANN}       {{\it ANN}}

\input{\PUBD/babarsym.tex}

\def\LUMI{211}  
\def\BBpairs{232}  
\def\BBpairsErr{3}  
\def\OffResLumi{22}  
\def\BRmean{6.9} 
\def\BRstat{2.0} 
\def\BRsyst{1.3} 
\def\BRsecmean{2.31} 
\def\BRsecstat{0.67} 
\def\BRsecsyst{0.42} 
\def\Acpmean{0.04} 
\def\Acpstat{0.29} 
\def\Acpsyst{0.05} 
\def\significance{3.6} 

\newcommand{\BaBarYear}      {05}
\newcommand{\BaBarNumber}    {006}
\newcommand{\SLACPubNumber} {11057}

\newcommand{\BaBarType}      {PUB}  

\begin{document}

\preprint{\babar-\BaBarType-\BaBarYear/\BaBarNumber}
\preprint{SLAC-PUB-\SLACPubNumber}

\title{Evidence for the Decay {\boldmath $\Btokstarpi$}}

%
\author{B.~Aubert}
\author{R.~Barate}
\author{D.~Boutigny}
\author{F.~Couderc}
\author{Y.~Karyotakis}
\author{J.~P.~Lees}
\author{V.~Poireau}
\author{V.~Tisserand}
\author{A.~Zghiche}
\affiliation{Laboratoire de Physique des Particules, F-74941 Annecy-le-Vieux, France }
\author{E.~Grauges}
\affiliation{IFAE, Universitat Autonoma de Barcelona, E-08193 Bellaterra, Barcelona, Spain }
% ...
\collaboration{The \babar\ Collaboration}
\noaffiliation

\date{April 6, 2005}

\begin{abstract}
We
have measured
the process \Btokstarpitokpipi\ with
\BBpairs\ million $\FourS\to\BB$ decays collected
with the \babar\ detector at the \pep2\ asymmetric-energy \BF\ at SLAC.
From a signal yield of $89\pm26$ events
we obtain the branching fraction
$\BR(\Btokstarpi)=[\BRmean \pm {\BRstat} (stat) \pm {\BRsyst} (syst)] \times 10^{-6}$
with a statistical significance of
\significance\ standard deviations
including systematic uncertainties,
and a charge asymmetry of
$\Acpmean \pm {\Acpstat} (stat) \pm {\Acpsyst} (syst)$.
\end{abstract}

\maketitle

%
%
%
%

Branching fraction and \CP-asymmetry measurements of charmless \B-meson decays
provide valuable constraints for the determination of the unitarity triangle
constructed from elements of the Cabibbo-Kobayashi-Maskawa quark-mixing
matrix~\cite{Cabibbo,KobayashiMaskawa}.
They test the accuracy of theoretical models such as those based on QCD
factorization~\cite{BenekeNeubert2003} or SU(3) flavor symmetry~\cite{ChiangGronau2004}.
It has been argued that the influence of final-state interactions
like charming penguins~\cite{Ciuchini1997,Ciuchini2001,Isola2003}
and similar long-distance rescattering effects~\cite{AtwoodSoni1998}
on both the branching fraction and \CP asymmetry of \Btokpi\  decays may be significant.
In this context, the decay \Btokstarpi\ is particularly interesting in the light of recent
measurements of direct \CP-violation in the $\Bz\to\K^\pm\pi^\mp$ and
$\B^\pm\to\K^\pm\pi^0$ channels~\cite{Bkpi2004,BelleK+pi0,BaBarK+pi0}.
Comparison to the $\Bpm\to\K^{*0}\pi^\pm$ decay mode~\cite{belleKstar0pi} can provide information
about the dominance of penguin diagrams.
Here we present a measurement of the branching fraction $\BR(\Bpm\to\Kstar(892)^\pm\piz)$
and its charge asymmetry
$$\Acp=\frac{N(\Bm\to\Kstarm\piz)-N(\Bp\to\Kstarp\piz)}{N(\Bm\to\Kstarm\piz)+N(\Bp\to\Kstarp\piz)}$$
based exclusively on \Bpm\ decays to the \Kpipi\ final state. 
%
%
%
The data used in this analysis were collected with the \babar\ detector~\cite{BabarDet}
at the \pep2\
asymmetric-energy \epem\ storage ring
at SLAC.
Charged-particle trajectories are measured by
a five-layer double-sided silicon vertex tracker and
a 40-layer drift chamber
located within a 1.5-T solenoidal magnetic field.
Charged hadrons are identified by combining energy-loss
information from tracking with the measurements from 
a ring-imaging Cherenkov detector.
Photons are detected by
a CsI(Tl) crystal electromagnetic calorimeter
with an energy resolution of $\sigma_E/E=0.023(E/\GeV)^{-1/4}\oplus 0.014$.
The magnet's flux return is instrumented for muon and \KL\ identification.

The data sample
includes $\BBpairs\pm\BBpairsErr$~million \BB\ pairs
collected at the \FourS resonance,
corresponding to
an integrated
luminosity of \LUMI~\ifb.
It is assumed that
neutral and charged \B meson pairs
are produced in equal numbers~\cite{BBProdRatio}.
In addition, \OffResLumi~\ifb\ of data collected at
40~\MeV\ below the \FourS resonance mass were used for background studies.
%
 
%
%
%
%
We performed full detector Monte Carlo (MC) simulations
equivalent to 
460 \ifb\ of generic \BB decays and 140 \ifb\ of continuum
quark-antiquark production events.
In addition, we simulated
over 30 exclusive charmless \B decay modes,
including 1.2 million signal \Btokstarpi\ decays.

\B meson candidates are reconstructed from one
charged track and two neutral pions.

%
%
%
%
The charged track used to form the \Btokstarpi\ candidate is required to
have at least 12 hits in the drift chamber, to have a transverse momentum greater
than 0.1~\GeVp, and to be consistent with originating from
a \B-meson decay.
Its signal in the tracking and Cherenkov detectors is required to be
consistent with that of a kaon.
The kaon selection algorithm is 70--92\% efficient within the relevant momentum range,
with a misidentification rate of less than 7\%. 
We remove tracks that pass
electron selection criteria based on $dE/dx$ and calorimeter information.
Neutral pion candidates are formed from two photons, each with a minimum energy
of 0.03~\GeV\ and a lateral moment~\cite{LATdef}
of their shower energy deposition
greater than zero and less than 0.6.
The angular acceptance of photons is restricted to exclude
parts of the calorimeter where showers are
not fully contained.
We require the photon clusters forming the \piz\ to be separated in space,
with a \piz\ energy of at least 0.2~\gev
and an invariant mass between 0.10 and 0.16~\GeVm.

Two kinematic variables,
$ \Delta E = E^{*}_{B} - \sqrt{s}/2$ and
the beam energy substituted mass
$  \mes = \sqrt{(s/2 + {\bf p}_{0}\cdot {\bf p}_{B})^{2}/E^{2}_{0} - {\bf p}^{2}_{B}}$,
are used for the final selection of events.
Here $E^{*}_{B}$ is the $\B$-meson-candidate energy in the center-of-mass frame,
$E_{0}$ and $\sqrt{s}$ are the total energies of the $\epem$ system in
the laboratory and center-of-mass frames, respectively, and ${\bf p}_{0}$ and
${\bf p}_{B}$ are the three-momenta of the $\epem$ system and the $\B$
candidate in the laboratory frame.
For correctly reconstructed \Kstarpi\ candidates \DeltaE\ peaks at zero, while
final states
with a charged pion,
such as $\Btorhopi$,
shift
$\DeltaE$
by approximately 80~\MeV\ on average.
Events are selected with
$5.20<\mes<5.29~\GeVm$  and
$|\DeltaE|<0.20~\GeV$.
The \DeltaE\ limits help remove background from two- and four-body \B decays
at a small cost to signal efficiency.

Continuum quark-antiquark production is the dominant background.
To suppress it,
we select only those events where the angle \thetabsph\
in the center-of-mass frame
between the
direction of the \B-meson-candidate
and the
sphericity axis of the rest of the event
satisfies  $|\cos \thetabsph| < 0.9$.
In addition, we construct a non-linear
discriminant,  implemented as an artificial neural network (\ANN)
that uses three input parameters: 
the zeroth- and second-order
Legendre event shape polynomials $L_0,L_2$ of the momenta and polar angles
of all candidates
in the rest of the event, and the output of a
multivariate, non-linear \B-meson-candidate tagging algorithm~\cite{BTagRef}.
\ANN\ is peaked at 0.5
for continuum-like events and at 1.0 for \B decays.
We require $\ANN>0.58$ for our event selection.
To further improve the signal-to-background ratio we
restrict
the effective invariant mass of the $\Kstar$ candidate
to $0.8 < m_{K\pi} < 1.0~\GeVm$.
Neutral-pion combinatorics lead to
$30\%$ of our signal events having more than
one candidate per event.
We choose the best candidate based on a \chisq\ formed from
the measured masses of the two \pizero\ candidates within the
event compared to the known \pizero\ mass~\cite{pdg}.

After the selection described above,
the \Btokstarpi\ selection efficiency is $16.5\%$.
In MC studies, the signal candidate is correctly reconstructed $(64.5\pm 6.5)\%$ of the time.
The remaining candidates come from
self-cross-feed (SCF) events,
which
stem primarily
from swapping the low energy \piz from the resonance with another from the rest of the event.
The fraction of SCF events in which the track was swapped with an oppositely charged track
was found to be negligible.

%
%
%
%

MC events are used to study backgrounds from other \B-meson decays.
The dominant
contribution comes from $b \rightarrow c$ transitions;
the next most important is from
charmless \B-meson decays.
The latter tend to be more problematic as the branching fractions
are often poorly known, and because they may peak at the same
invariant mass as the
signal \Btokstarpi\ events.
Thirteen individual charmless modes show a significant contribution
once
the event selection has been
applied.
These modes are added into the fit fixed at the yield and asymmetry
determined by the simulation.
Wherever branching fractions are not available, we use half
the upper limit. If no charge asymmetry measurement is available,
we assume zero asymmetry.

Although all other known resonant \Kstar\ states
-- subsequently referred to as \Kstarstar\ --
lie outside our $\Kstar(892)$ mass window, some may still contribute due
to their large width.
To estimate the contribution to the signal we select a region in the
\Kpi\ invariant mass between 1.2 and 1.6 \gevcc, 
motivated by the presence of the broad $\Kstar_{0}(1430)$ resonance
which decays predominantly to $K\pi$.
In this region we make a full maximum likelihood
fit to the three variables \DeltaE, \mes\ and \ANN\ in an
analogous way to how we fit our signal (see below), and 
extrapolate the result of this fit to the $\Kstar(892)$ signal region
using a \Bpm\to $K^*_0(1430)^{\pm}$\piz MC.
The fit to the $\Kstar_{0}(1430)$ region yields $263 \pm 34$ events,
which translates to 34 \Kstarstar\ events contributing
to the background in our signal region.
We assign a 100\% systematic uncertainty to this number
to cover possible interference effects as well as uncertainty arising from
the lineshapes of \Kstarstar\ resonances, which are not well established.

The non-resonant \Btokpipi\ branching fraction has, to date, not been
measured.
To estimate the significance of its
contribution we select a region of the Dalitz plot of $m^2_{K\pi}$
-- for the \piz\ from primary and secondary decay -- 
that is far
from the signal as well as $\Kstar(1430)$ and higher \kstarstar\ resonances and
which has low levels of continuum background.
A likelihood fit in this region yields
$6\pm8$
events,
which translates into less than three events in our \Kstar\ signal region,
assuming the non-resonant events are distributed evenly across the Dalitz plot. 
We consequently deem the non-resonant contribution negligible.

%
%
%
%

An unbinned maximum likelihood fit to the variables
\mes, \DeltaE, $m_{K\pi}$, and \ANN\
is used to extract
the total number of signal \Btokstarpi\ and continuum background
events and their respective charge asymmetries.
The likelihood for the selected sample is given by the product
of the probability density functions (PDF) for each individual candidate,
multiplied by the Poisson factor:
$$    \Like = \frac{1}{N!}\,e^{-N^\prime}\,(N^\prime)^N\,\prod_{i=1}^N {\cal P}_i\, ,$$
where $N$ and $N^\prime$ are the number of observed and expected events, respectively.
The PDF ${\cal P}_i$ 
for a given event
$i$ is the sum of the signal and background terms:
\begin{eqnarray}
  {\cal P}_i
  & = &
  N^{\rm Sig} \times  \frac{1}{2} \,
  {\large[}\,(1 - Q_i A^{\rm Sig})\, f\, {\cal P}^{\rm Sig}_{{\rm SCF},i}     
   \nonumber
  \\
  &  &
    ~+~ (1-Q_i A^{\rm Sig}) (1 - f)\, {\cal P}^{\rm Sig}_{i} 
  \,{\large]}
  \nonumber
  \\
  & & 
  + \sum _j N^{\rm Bkg}_{j}\times  \frac{1}{2}  (1 - Q_iA^{\rm Bkg}_{j})\,  {\cal
    P}^{\rm Bkg}_{j,i},
  \nonumber 
  \label{pdfsum}
\end{eqnarray}
where $Q_i$ is the charge of the kaon in the event,
$N^{\rm Sig}(N^{\rm Bkg}_j)$ and $A^{\rm Sig}(A^{\rm Bkg}_j)$ are the yield and asymmetry for
signal and background component $j$, respectively, 
and $f=35.5\%$ is the
fraction of SCF signal events.
The $j$ individual background terms comprise continuum, $b\to c$ decays, \Kstarstar,  and
13 exclusive charmless \B decay modes.
The PDF for
each component, in turn, is the product of the PDFs for each of the
fit input variables,
$  {\cal P} = {\cal P}_{{\rm m_{ES}},\Delta E}{\cal P}_{\ANN}{\cal P}_{m_{K\pi}}.$
Due to correlations between \DeltaE\ and \mes,
the ${\cal P}_{{\rm m_{ES}},\Delta E}$  for signal
and all background from \B decays are described by two-dimensional
non-parametric PDFs~\cite{2dkeys} obtained from MC.
For continuum background, the correlations in \DeltaE\ and \mes\ are
$\sim 1\%$, hence a separate PDF is used for each of them;
\mes\ is well described by
an empirical phase-space threshold
function~\cite{argus} and
\DeltaE\ is parameterized with a second degree polynomial.
The parameters of the continuum PDFs are floated in the fit.
$\ANN$ is described by a non-parametric PDF for continuum
background and by a Crystal Ball function~\cite{CrystalBall} for all other modes.
For ${\cal P}_{m_{K\pi}}$, one-dimensional
non-parametric PDFs obtained from MC
are used to describe all
modes except the signal mode itself, which is described by
a Breit-Wigner line-shape combined with a first degree polynomial.
The parameters for this PDF are held fixed to the MC values and varied within
errors to estimate systematic uncertainties.

A number of cross checks confirm that the fit is unbiased.
In 1000 separate MC experiments
we generate the expected number of events for the various fit
components before using the maximum likelihood fit to extract the yields and
asymmetries.
The distributions for each component are generated from the component's PDF,
giving values for the fit variables \mes, \DeltaE, \ANN, and $m_{K\pi}$.
The expected number of events is calculated from the branching fraction and
efficiency for each individual mode.
The generated number of events for each fit component is determined by fluctuating
the expected number according to a Poisson distribution.
The test is repeated using
samples with differing asymmetry values.
We repeat these MC studies
using
fully simulated
signal \Btokstarpi\
events 
instead of generating the signal component from our PDFs.
This verifies that the signal component is correctly modeled
including correlations between the fit variables.
Finally, omitting $m_{K\pi}$ as a fit variable
has no significant influence on the signal yield,
indicating that our treatment of \Kstarstar\ background
is indeed effective.

%
%
%

\renewcommand\baselinestretch{1.3}
\begin{table}[bt]
  \caption[Systematic uncertainties]{Breakdown of systematic uncertainties.}
  \begin{ruledtabular}
    \begin{tabular}{ll}
      \multicolumn{2}{c}{Absolute systematic uncertainties on yields}\\ 
      Source                   &  $\sigma_{\rm Syst.}^{\rm Yield}$ (Events)\\
      \hline
      Background normalization & $\pm 14$ \\ 
      PDF shapes               & $^{+\;\, 2.1}_{-\;\, 4.0}$ \\ 
      SCF fraction             &  $\pm\;\, 1.8$ \\ 
      $\Delta$E shift          &  $\pm\;\, 2.2$  \\ 
      \hline
      Total                    &  $\pm 14$ \\
      \hline\hline
      \multicolumn{2}{c}{Relative systematic uncertainties on \BR(\Btokstarpi) }\\
      Source            &  $\sigma_{\rm Syst.}^{\BR} (\%) $ \\
      \hline
      Efficiency estimation        &  $\pm 7.3 $ \\ 
      \B\ counting                 &  $\pm 1.1 $ \\
      \hline
      Total      & $\pm 7.4$ \\
      \hline\hline
      \multicolumn{2}{c}{Systematic uncertainties on \Acp }\\
      Source            &  $\sigma_{\rm Syst.}^{\Acp }$ \\
      \hline
      Background normalization      &  $^{+0.018}_{-0.010}$ \\ 
      Detector asymmetry            &  $\pm 0.003 $ \\ 
      Background asymmetry          &  $^{+0.049}_{-0.041}$ \\ 
      \hline
      Total                         &  $^{+0.054}_{-0.043}$  \\
    \end{tabular}
  \end{ruledtabular}
  \label{tab:systematics}
\end{table}
\renewcommand\baselinestretch{1.0}

Individual contributions to the systematic uncertainty are
summarized in Table \ref{tab:systematics}.
We calculate the uncertainty of the continuum background estimation
directly from the fit to data.
The
backgrounds from \B decays are determined from simulation and fixed
according to
their efficiencies and branching fractions.
For those individual decay modes which have been measured
we vary the number of events in the fit by 
their measured uncertainty.
For all others we vary the amount included in
the fit by $\pm 100\%$. For the
$b\to c$
component we fix the rate
based on the number calculated from MC samples and vary the
amount based on the statistical uncertainty of this number (6\%).
The shifts in the fitted yields are calculated
for each mode in turn and then added in quadrature to find the total
systematic effect. The largest individual contribution comes from the
\Kstarstar\ estimation.

%
%
\begin{figure}[bt]
  \renewcommand\baselinestretch{0.5}
  \begin{tabular}{cc}
    \epsfig{file=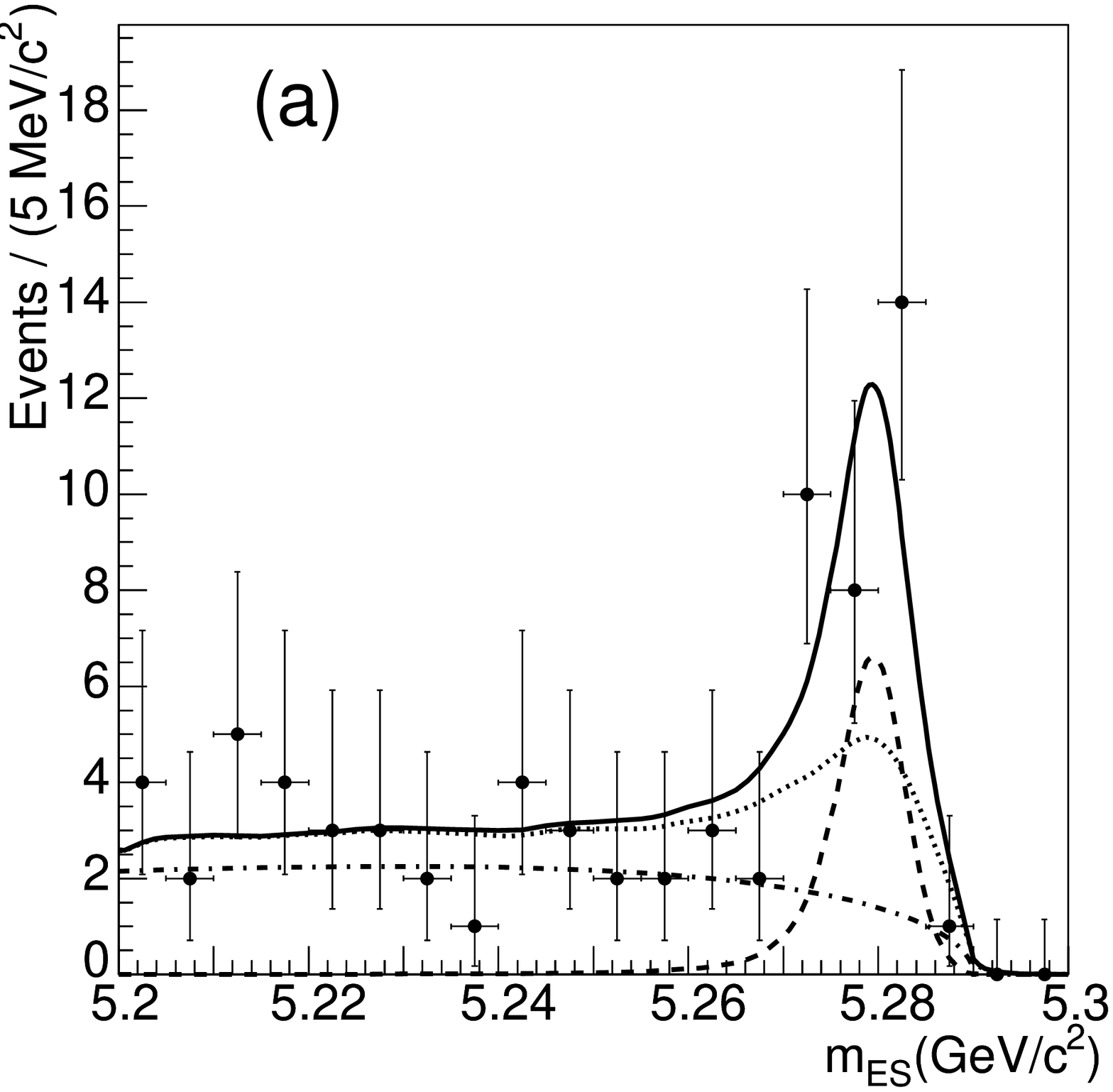,width=0.52\columnwidth}
    &
    \epsfig{file=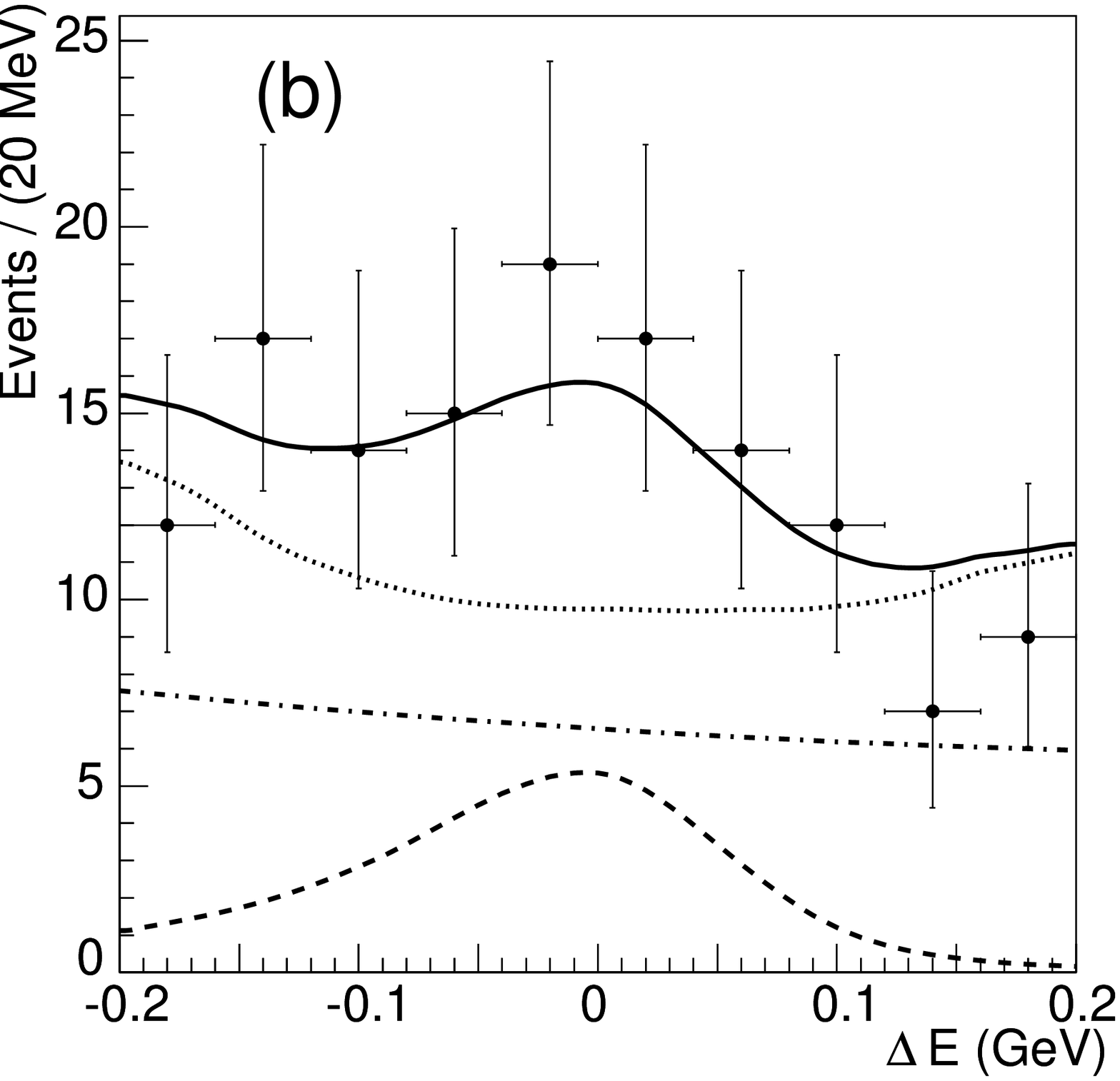,width=0.52\columnwidth}
    \\
    \epsfig{file=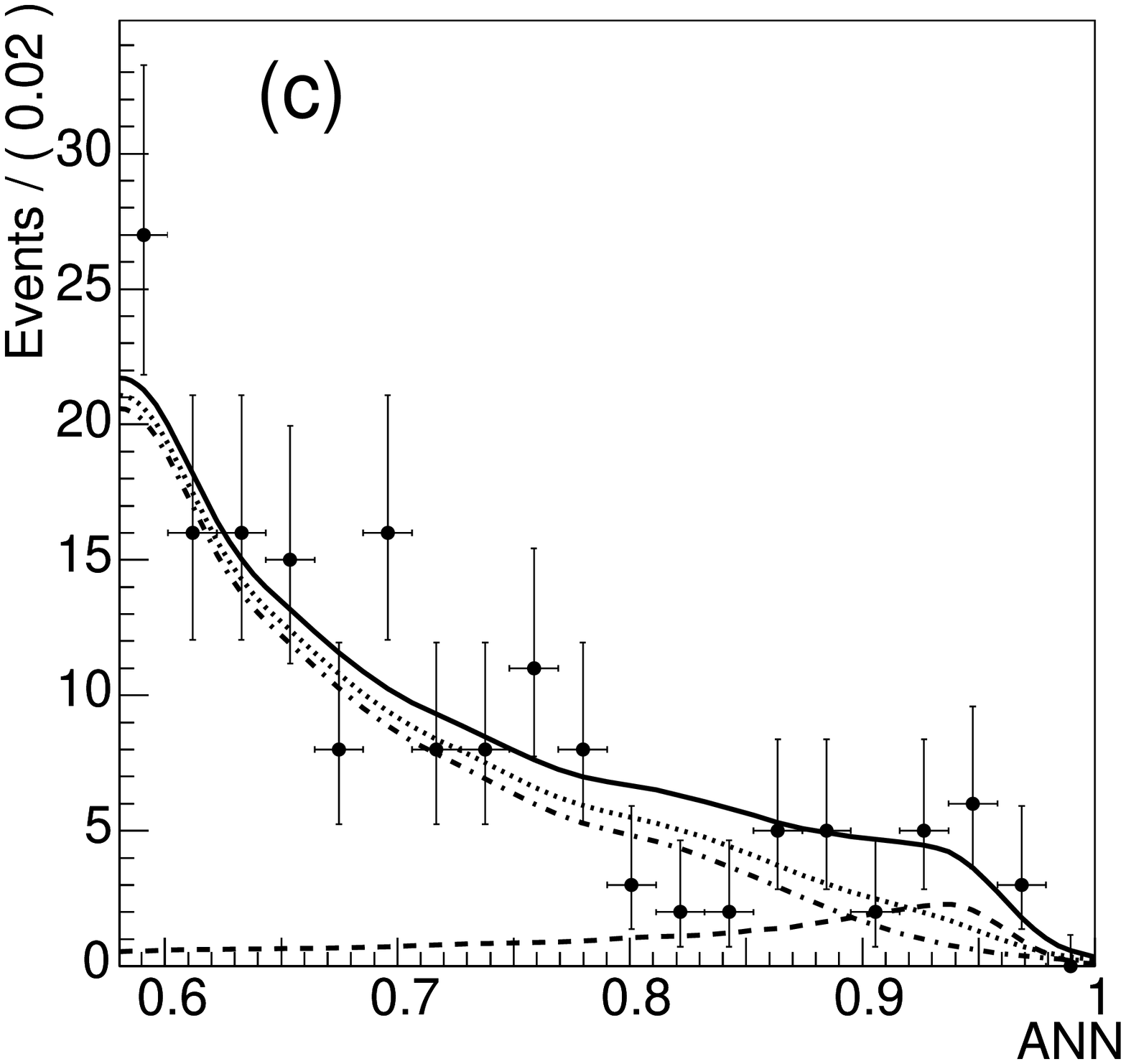,width=0.53\columnwidth}
    &
    \epsfig{file=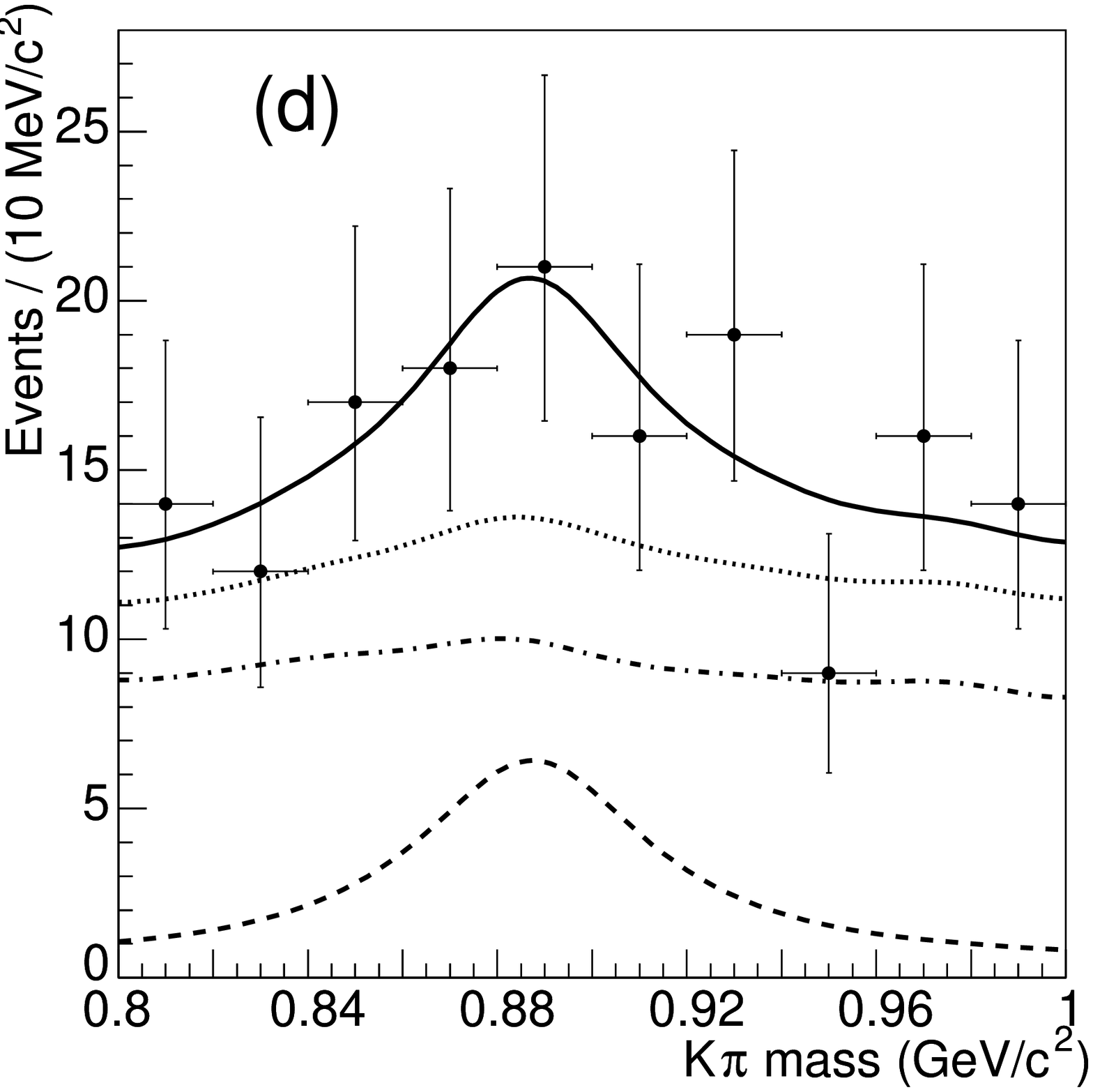,width=0.52\columnwidth}
    \\
  \end{tabular}
  \renewcommand\baselinestretch{1.0}
  \caption[Results for the four variables of the maximum likelihood fit]
  {Likelihood projection plots for the four fit variables,
    (a) \mes, (b) \DeltaE, (c) \ANN, and (d) $m_{K\pi}$.
    In each plot the solid line represents the total PDF,
    the dotted line represents the total background,
    the dotted-dashed line represents the continuum contribution,
    and the dashed line represents the signal component.
  The plots contain a subset of the events defined by a likelihood ratio
  of at least 0.1 (see text).}
  \label{fig:results}
\end{figure}
%

To take into account the variation of the two-dimensional
non-parametric PDFs used for
\DeltaE and \mes,
we smoothen the MC-generated distributions from which
the PDFs are derived.
For
$m_{K\pi}$ and $\ANN$,
the parameterizations determined
from fits to MC events are varied by one standard deviation. 
The systematic uncertainties are determined using the altered PDFs and
fitting to the final data sample. The overall shifts in the central
value are taken as the size of the systematic uncertainty.

We vary the SCF fraction by
a conservative estimate of its relative uncertainty
($\pm10\%$) and assign the shift
in the fitted number of signal events as the systematic uncertainty of the SCF
fraction.

To account for differences in the neutral-particle reconstruction
between data and MC simulation
the signal PDF distribution in $\Delta$E is offset by $\pm 5 \mev$ and the data refitted. 
The larger of the two shifts in the central value of the yield is 2.2 events, which is taken
as the systematic uncertainty for this effect.

Corrections to the \pizero\ energy distribution,
determined using various control samples,
add a systematic uncertainty
of 7.2\%.
A relative systematic uncertainty of $1\%$ is assumed
for the
kaon identification.  
A relative systematic uncertainty of 0.8\% on the efficiency for a single charged
track is applied.
Adding all the above contributions in quadrature gives a
relative systematic uncertainty on \BR\ of 7.3\%.
Another contribution of $1.1\%$ comes from
the uncertainty on the
total number of \B events. 

The cross section for the interaction of kaons with protons and
neutrons differs with charge. At low momenta this
can introduce a bias to the observed charge asymmetry.
We estimate this bias by modelling the
average loss of kaons from a sample based on the \Kstarpi\ signal MC
using the known detector material
constants, and
find ${\cal A}_{Kp}= -0.0031 \pm 0.0006$, which is
negligible compared to the precision at which we measure \Acp.

To calculate the effects of systematic
shifts in the charge asymmetries of background modes, each
mode is varied by
its
measured uncertainty.
For contributions with no measurement, we assume zero asymmetry and
assign an uncertainty of 20\%, motivated by the largest charge asymmetry
measured in any mode so far~\cite{Bkpi2004}.
The individual shifts are then added in
quadrature to find the total systematic uncertainty. The greatest individual
contribution comes from the \Kstarstar\ estimate.
In addition, the effect
of altering the normalizations of the \B backgrounds affects the
fitted asymmetry.
The size of the shift on the fitted \Acp\ is taken as the size of
the systematic uncertainty.
%

%
%

A total of 23,465 events were fitted, of which 11,960 had positively charged candidates.
The central value of the signal yield from the maximum likelihood fit is
$89\pm26$ events, over
an expected background of $634\pm 40$ events from other \B decays.
We obtain a branching fraction of
$\BR(\Btokstarpitokpipi)
= [\BRsecmean \pm {\BRsecstat} \pm {\BRsecsyst}] \times 10^{-6}$
and charge asymmetry of
$ \Acp(\Btokstarpitokpipi)
= \Acpmean \pm {\Acpstat} \pm {\Acpsyst},$
where the first error is statistical and the second one systematic.
Alternatively, we calculate the 90\% confidence upper limit on the
\Btokstarpitokpipi\ branching fraction to be $3.9\times 10^{-6}$.
Compared against the null hypothesis, the statistical significance
$\sqrt{-2 \ln (\Like_{Null}/\Like_{Max})}$ 
of the yield amounts to 4.1 standard deviations.
The fit was redone fixing the signal yield to the
lowest yield allowed accounting for all possible combinations
of systematic uncertainties.
The significance of this result corresponds to \significance\ standard deviations.
 
The results of the fit are illustrated in 
Fig.~\ref{fig:results}.
The plots are enhanced in signal
by selecting only those events which exceed a threshold of 0.1 for the likelihood ratio
$R = (N^{\rm Sig}{\cal P}^{\rm Sig})/(N^{\rm Sig}{\cal P}^{\rm Sig} + \sum_{i} N^{\rm Bkg}_{i}{\cal P}^{\rm Bkg}_{i})$,
where $N$ are the central values of the yields from the fit and
${\cal P}$ are the PDFs with the projected variable integrated out.
This threshold is optimized by maximizing the ratio
$S = (N^{\rm Sig}~\epsilon^{\rm Sig})/(\sqrt{N^{\rm Sig}~\epsilon^{\rm Sig} + \sum_{i} N^{\rm Bkg}_{i}~\epsilon^{\rm Bkg}_{i}})$
where $\epsilon$ are the efficiencies after the threshold is applied.
The PDF components are then
scaled by the appropriate $\epsilon$.

In conclusion, we have measured
the charge asymmetry and branching fraction for the
decay \Btokstarpitokpipi\ using
a maximum likelihood fit.
Assuming a secondary branching fraction of 1/3 for the \Kstartokpi\ final state
our result implies
$\BR(\Btokstarpi) = [\BRmean \pm \BRstat \pm \BRsyst]\times 10^{-6}$, and
a charge asymmetry of
$ \Acp = \Acpmean \pm {\Acpstat} \pm {\Acpsyst}$
where the first error is statistical and the second error systematic.
The statistical significance of the branching fraction result
including systematic uncertainties
is calculated to be \significance\ standard deviations, showing evidence for this decay.
The systematic error of the branching fraction and asymmetry is dominated by
the contribution of \Kstarstar\ resonances.

%
%

We are grateful for the excellent luminosity and machine conditions
provided by our \pep2\ colleagues, 
and for the substantial dedicated effort from
the computing organizations that support \babar.
The collaborating institutions wish to thank 
SLAC for its support and kind hospitality. 
This work is supported by
DOE
and NSF (USA),
NSERC (Canada),
IHEP (China),
CEA and
CNRS-IN2P3
(France),
BMBF and DFG
(Germany),
INFN (Italy),
FOM (The Netherlands),
NFR (Norway),
MIST (Russia), and
PPARC (United Kingdom). 
Individuals have received support from CONACyT (Mexico), A.~P.~Sloan Foundation, 
Research Corporation,
and Alexander von Humboldt Foundation.

\bibliography{KstPi0_References}
\addcontentsline{toc}{section}{\numberline{}References}
\bibliographystyle{apsrev}

\end{document}
%